\documentclass[12pt]{iopart}
\usepackage{amssymb,epsfig}
\newcommand{\nab}[2]{\nabla_{\displaystyle{#1}}#2}

\newcommand{\Lie}[2]{{\mathcal{L}}_{\displaystyle{#1}}#2}
\newcommand{\Riem}{\mathrm{Riem}}
\newcommand{\de}{\,\mathrm d}
\newcommand{\Par}[2]{\frac{\partial{#1}}{\partial{#2}}}
\newcommand{\ha}{{\textstyle{\frac{1}{2}}}}

\begin{document}

\title{How to Lasso a Plane Gravitational Wave}
\author{P.C. Bollada}

\address{}

\begin{abstract}
Beginning with the stress-energy tensor of an elastic string this
paper derives a relativistic string and its form in a parallel
transported Fermi frame including its reduction in the Newtonian
limit to a Cosserat string. In a Fermi frame gravitational
curvature is seen to induce three dominant relative acceleration
terms dependent on: position, velocity and position, strain and
position, respectively. An example of a string arranged in an
axially flowing ring (a lasso) is shown to have a set of natural
frequencies that can be parametrically excited by a monochromatic
plane gravitational wave. The lasso also exhibits, in common with
spinning particles, oscillation about geodesic motion in
proportion to spin magnitude and wave amplitude when the spin axis
lies in the gravitational wave front. Coordinate free notation is
used throughout including the development of the properties of the
Fermi frame.
\end{abstract}



\section{Introduction}
\label{Intro}
The use of an elastic continuum to detect gravitational effects is
not new \cite{MTW}, but a space based gravitational wave detector
would have to be of considerable size to resonate with the low
frequency and strength of waves theoretically predicted
\cite{Finn}. The use of a one dimensional elastic continuum holds
out the possibility of producing such a resonant detector without
the burden of high payload for shipment to space. A one
dimensional elastic continuum is known as a Cosserat string and is
the limit of a Cosserat rod as the cross sectional area tends to
zero. The theory can be consulted in \cite{Ant} and is
fundamentally formulated in the Lagrangian picture in which each
element of the string is labeled by $s\in[0,L_0]$ where $L_0$ is
the string's reference (unstretched) length. The evolution of the
string at time $t$ is given by a space curve
\begin{eqnarray*}
s\in[0,L_0]\mapsto  {\bf R}(s,t)
\end{eqnarray*}
and obeys the Cosserat string equation
\begin{eqnarray}
\rho A\partial_t^2{\bf R}=\partial_s{\bf N}+{\bf F}
\label{10_Master_Equation}
\end{eqnarray}
where in general the mass density
\begin{eqnarray*}
s\in[0,L_0]\mapsto \rho(s)
\end{eqnarray*}
and area
\begin{eqnarray*}
A\in[0,L_0]\mapsto A(s)
\end{eqnarray*}
The mass density, $\rho$, and $A$ are assumed constant in this
paper allowing $A$ to be removed and $\rho$ to be reinterpreted as
mass per unit length . To close the equations
(\ref{10_Master_Equation}) the contact force, ${\bf N}$ and
external force ${\bf F}$ are prescribed. In the string model the
contact force ${\bf N}$ is aligned along the tangent to the
space-curve. A simple constitutive relation due to Kirchhoff is
given by
\begin{eqnarray}
{\bf N}\equiv EA\left(|{\bf R}'|-1\right)\frac{{\bf R'}}{|{\bf
R}'|} \label{Constitutive}
\end{eqnarray}
where $E$ is Young's modulus. The Kirchhoff constitutive relation
is linear and is a good model for small strains for strings in
permanent tension. Generalizations are given in \cite{Ant} and an
application for rods in \cite{TuckerWang}.

The dynamics of a Cosserat string in a Newtonian spherically
symmetric gravitational field is modeled by setting
\begin{eqnarray*}
{\bf F}\equiv-\frac{GM\rho A{\bf R}}{|{\bf R}|^3}.
\end{eqnarray*}
To model the stresses on a Cosserat string due to spacetime
curvature, for example, in a gravitational wave, or in orbit about
a black hole, one can supply a `tidal' force lifted from general
relativity:
\begin{eqnarray*}
{\bf F}\equiv \rho A\Riem(\dot C,{\bf R},\dot C,-),
\end{eqnarray*}
where $\dot C$ is the tangent to the world line of a comoving
observer and $Riem$ is the curvature tensor \cite{Tucker}. The
tidal term represents the relative acceleration of two \emph{non
interacting} neighbouring particles and it is not clear how this
term should be modified when particles are not in free fall and
are interacting. This subject is addressed in this paper.

\section{The relativistic string}
\subsection{Stress-energy tensor for a string}
For a general spacetime $\cal M$ with metric $g$, the world sheet
of a string can be represented by two orthonormal vector fields
$V$ and $W$ such that $g(V,V)=-1$, $g(W,W)=1$ and $g(V,W)=0$. We
associate the integral curves of $V$ with material points on the
string. The $(2,0)$ stress-energy tensor for the string can be
written
\begin{eqnarray*}
T\equiv \rho\tilde V\otimes\tilde V+p\tilde W\otimes\tilde W
\end{eqnarray*}
where the metric dual $\tilde V\equiv g(V,-)$, $\rho$ is the
density  and $p$ is the pressure  at points on the string. For a
string it is convenient to interpret $\rho$ as mass per unit
length and $p$ as force.
To find the equation for a relativistic string we wish to set the
divergence $\nabla\cdot T\equiv(\nab{X_a}{T})(-,X^a)$ to zero,
where $\nabla$ is compatible with the metric $g$ in the absence of
the string and $\{X_a\}$ is some basis. However, $\nabla\cdot T$
is not defined in directions out of the tangent plane of the
string world sheet. To this end denoting an orthonormal frame
$\{X_a\},a\in[0,3]$ such that $X_0=V,X_1=W$, we demand
$\nab{X_2}{T}=\nab{X_3}{T}=0$. Then
\begin{eqnarray}
0&=&\nabla\cdot T\equiv(\nab{X_a}{T})(-,X^a)\nonumber\\
 &=&-\nab{V}{T}(-,V)+\nab{W}{T}(-,W)\nonumber\\
 &=&\tilde V\nab{V}{\rho}+\rho\tilde V\nabla\cdot V+\rho\nab{V}{\tilde
V}+p\nab{W}{\tilde W}+\tilde W\nab{W}p+p\tilde W\nabla\cdot W
 \label{RS1}
\end{eqnarray}
where we use $\tilde V(\nab{V}{V})=\tilde W(\nab{W}{W})=0$ to
eliminate two terms. Also note that $\nabla\cdot V=\tilde W(
\nab{W}{V})=-\tilde V(\nab{W}{W})$ and $\nabla\cdot W=-\tilde
V(\nab{V}{W})=\tilde W(\nab{V}{V})$. Taking the metric dual of
(\ref{RS1}) gives the relativistic string equation
\begin{eqnarray}
 V\nab{V}{\rho}+\rho
V\nabla\cdot V+\rho\nab{V}{V}+p\nab{W}{W}+W\nab{W}{p}+
pW\nabla\cdot W=0. \label{RS}
\end{eqnarray}
\section{The Special relativistic Cosserat string} \label{SS2}

For a general spacetime ${\mathcal M}$ with metric $g$, a string
can be defined by the embedding $R$
\begin{eqnarray*}
\mathbb{R}\times [0,1]\rightarrow\mathcal M,\nonumber\\
(\tau,\sigma)\mapsto R(\tau,\sigma).
\end{eqnarray*}
Now let $(\mathcal M,g)$ be Minkowski spacetime  with natural
Cartesian coordinates $\{t,x,y,z\}$ and metric
\begin{eqnarray*}
g\equiv -\de t\otimes\de t+\de x\otimes\de x+\de y\otimes\de y+\de
z\otimes\de z
\end{eqnarray*}
then the string can be defined by the mapping
\begin{eqnarray*}
\mathbb{R}\times [0,1]\rightarrow\mathcal M,\nonumber\\
(\tau,\sigma)\mapsto R(\tau,\sigma)
 &=&[t=T(\tau,\sigma),{\bf R}(\tau,\sigma)]\\
 &\equiv&[t=T(\tau,\sigma),x=X(\tau,\sigma),y=Y(\tau,\sigma),z=Z(\tau,\sigma)],
\end{eqnarray*}
 Using notation $\dot{()}\equiv\partial/\partial\tau$ and
$()'\equiv
\partial/\partial\sigma$ we can write
\begin{eqnarray}
\partial_\tau=\dot T\partial_t+\dot{\bf R}
\label{A20}
\end{eqnarray}
and
\begin{eqnarray}
\partial_\sigma=T'\partial_t+{\bf R}'
\label{A30}
\end{eqnarray}
where
\begin{eqnarray*}
\dot{\bf R}\equiv\dot X\partial_x+\dot Y\partial_y+\dot
Z\partial_z
\end{eqnarray*}
\begin{eqnarray*}
{\bf R}'\equiv X'\partial_x+ Y'\partial_y+ Z'\partial_z.
\end{eqnarray*}
Then setting
\begin{eqnarray*}
V=\alpha(\dot T\partial_t+\dot{\bf R})
\end{eqnarray*}
gives $g(V,V)=-1$, with $\alpha$ defined
\begin{eqnarray*}
\alpha \equiv[-g(\partial_\tau,\partial_\tau)]^{-1/2}=[\dot
T^2-\dot{\bf R}\cdot\dot{\bf R}]^{-1/2}
\end{eqnarray*}
and $\dot T^2>\dot{\bf R}\cdot\dot{\bf R}$ for $V$ to be time
like. Similarly
\begin{eqnarray*}
W=\beta( T'\partial_t+{\bf R}')
\end{eqnarray*}
gives $g(W,W)=1$, with $\beta$ defined
\begin{eqnarray*}
\beta
 \equiv[g(\partial_\sigma,\partial_\sigma)]^{-1/2}=[{\bf R}'\cdot{\bf R}'-T'^2]^{-1/2}
\end{eqnarray*}
and we demand ${\bf R}'\cdot{\bf R}'>T'^2$.  We now have
\begin{eqnarray}
V\nab{V}{\rho}=\alpha^2\dot\rho(\dot T\partial_t+\dot{\bf R}),
 \label{S1}
\end{eqnarray}
\begin{eqnarray}
W\nab{W}{p}=\beta^2p'( T'\partial_t+{\bf R}'),
 \label{S2}
\end{eqnarray}
\begin{eqnarray}
\rho\nab{V}{V}=\rho\alpha[(\dot\alpha\dot T+\alpha\ddot
T)\partial_t+\dot\alpha\dot{\bf R}+\alpha\ddot{\bf R}],
 \label{S5}
\end{eqnarray}
\begin{eqnarray}
p\nab{W}{W}=p\beta[(\beta' T'+\beta T'')\partial_t+\beta'{\bf
R}'+\beta{\bf R}''].
 \label{S6}
\end{eqnarray}
\begin{eqnarray}
\rho V\nabla\cdot V
 &=&\rho V [g(W,\nab{W}{V})-g(V,\nab{V}{V})]\nonumber\\
 &=&\rho V g(W,\nab{W}{V})\nonumber\\
 &=&-\rho V g(V,\nab{W}{W})\nonumber\\
 &=&\rho\alpha^2\beta^2(T''\dot T-{\bf R''\cdot\dot R})(\dot
 T\partial_t+\dot{\bf R}),
 \label{S3}
\end{eqnarray}
\begin{eqnarray}
p W\nabla\cdot W
 &=&p W [-g(V,\nab{V}{W})+g(W,\nab{W}{W}]\nonumber\\
 &=&-p W g(V,\nab{V}{W})\nonumber\\
 &=&p W g(W,\nab{V}{V})\nonumber\\
 &=&p\alpha^2\beta^2({\bf \ddot R\cdot R'}-\ddot T T')(
 T'\partial_t+{\bf R}'),
 \label{S4}
\end{eqnarray}
Where the orthonormality condition,
$g(V,W)=0=g(\partial_\tau,\partial_\sigma)$, implying
\begin{eqnarray}
\dot T T'={\bf \dot R R'},
 \label{Orth}
\end{eqnarray}
has been used to simplify  equations (\ref{S3}) and (\ref{S4}).
Further simplification comes from noting that, (\ref{Orth}) also
implies
\begin{eqnarray*}
(\dot TT')'=\dot T'T'+\dot T T''={\bf \dot R'\cdot R'+\dot R\cdot
R''}
\end{eqnarray*}
so that
\begin{eqnarray*}
T''\dot T-{\bf R''\cdot\dot R}&=&{\bf\dot R'\cdot R'}-\dot T'T'\\
  &=&\ha\Par{}{\tau}\left({\bf R'\cdot
R'}-(T')^2\right)\\
 &=&\ha\Par{}{\tau}\left(\frac{1}{\beta^2}\right)=-\frac{\dot\beta}{\beta^3}
\end{eqnarray*}
Defining $\rho_0\equiv\rho/\beta$ and adding equations (\ref{S1}),
(\ref{S3}) and (\ref{S5}) yields
\begin{eqnarray}
V\nab{V}{\rho}+\rho V\nabla\cdot
V+\rho\nab{V}{V}=\alpha\beta[\partial_\tau(\alpha\dot
T\rho_0)\partial_t+\partial_\tau(\alpha\rho_0\dot{\bf R})].
 \label{S7}
\end{eqnarray}
Similarly
\begin{eqnarray*}
\Par{}{\tau}(\dot T T')&=&\ddot T T'+\dot T\dot T'={\bf \ddot
R\cdot R'+\dot R\cdot \dot R'}
\end{eqnarray*}
implies
\begin{eqnarray*}
{\bf \ddot R\cdot R'}-\ddot T T'&=&\dot T\dot T'-{\bf \dot
R'\cdot\dot R}\\
 &=&\ha(\dot T^2-{\bf \dot R\cdot\dot R})'\\
 &=&\ha\left(\frac{1}{\alpha^2}\right)'=-\frac{\alpha'}{\alpha^3}.
\end{eqnarray*}
Then a similar simplification can be made by defining $p_0\equiv
p/\alpha$, which gives on adding  (\ref{S2}),(\ref{S4}) and
(\ref{S6})
\begin{eqnarray}
W\nab{W}{p}+p W\nabla\cdot
W+p\nab{W}{W}=\alpha\beta\left[\left(\frac{\beta p_0}{\dot
T}\,{\bf \dot R\cdot R'}\right)'\partial_t+(\beta p_0{\bf
R}')'\right]
 \label{S8}
\end{eqnarray}
Equations (\ref{S7}) and (\ref{S8}) imply both
\begin{eqnarray}
\Par{}{\tau}(\alpha\dot T\rho_0)+\left( \frac{\beta p_0}{\dot
T}\,{\bf \dot R\cdot R'}\right)'=0
 \label{S9}
\end{eqnarray}
and
\begin{eqnarray}
\Par{}{\tau}(\alpha\rho_0\dot{\bf R})+(\beta p_0{\bf R}')'=0.
 \label{S10}
\end{eqnarray}
Note, the familiar quantities: velocity, ${\bf v}\equiv\dot{\bf
R}/\dot T$; speed $v\equiv\sqrt{{\bf v\cdot v}}$ and Lorentz
factor $\gamma\equiv(1-v^2)^{-1/2}=\dot T\alpha$, can be used to
aid physical interpretation and clean up the appearance of
(\ref{S9}) and (\ref{S10}) still further. To complete the
relativistic string in Minkowski spacetime there is freedom to
relate $T$ on a world line corresponding to one material point to
the proper time of some observer, e.g. we could choose
$t=T(\tau,\sigma_0)=\tau$ for some $\sigma_0\in[0,1]$, implying
$\dot T(\tau,\sigma_0)=1$. This choice guarantees for general
points at non-relativistic speeds, $v\rightarrow 0$, that
$t=T(\tau,\sigma)\rightarrow\tau$, $\alpha\rightarrow 1$ and
$\beta\rightarrow 1/|{\bf R}'|$.

Given constitutive relations for $\rho_0$ and $p_0$ the special
relativistic equations (\ref{S9}) and (\ref{S10}) are 4 equations
for the unknowns $T(\tau,\sigma)$ and ${\bf R}(\tau.\sigma)$.
Setting $\rho_0$ to be independent of $\tau$ may be physically
interpreted as the string having no fluid properties, i.e. purely
elastic. However, in the relativistic context here, $\rho_0$
includes energy density with a time dependent contribution, for
example, due to heating and elastic potential. Choosing $\rho_0$
independent of $\tau$ is necessary to the reduction in the
Newtonian limit to the Cosserat string. With this choice
(\ref{S10}) becomes in the Newtonian limit:
\begin{eqnarray}
\rho_0\ddot{\bf R}={\bf N}'
 \label{A60}
\end{eqnarray}
where ${\bf N}\equiv -\beta p_0{\bf R'}\rightarrow -p_0{\bf
R'/|R'|}$.
Equation (\ref{A60}) is the Cosserat string
(\ref{10_Master_Equation})
 in the absence of external forces with
the notation reference mass per unit length, $\rho_0$, replacing
$\rho A$, the reference mass per unit volume times cross-sectional
area. We note that (\ref{S9}) may be written
\begin{eqnarray*}
\partial_\tau(\gamma\rho_0)=\left({\bf v\cdot N}\right)'
\end{eqnarray*}
which relates the rate of change of energy density to work
density.
\section{Fermi normal coordinates}
This section takes a geometric approach to the development of
Fermi coordinates for use in discussing the dynamics of the string
in a gravitational field as seen by a geodesic observer. For a
standard development of Fermi normal coordinates see
\cite{Nesterov}. A Fermi normal frame represents the coordinates
associated with the one-one exponential map between an observer's
tangent plane and the manifold in the observer's local
neighbourhood. An event local to the observer is given a time
component equal to the proper time of the observer and three
Cartesian like coordinates give the position. Let $\{x^a\},
a\in[0,3]$ denote these coordinates, where coordinate, $x^0$ is
the proper time of the observer. Henceforth only geodesic
observers will be considered. For an observer curve, $C$, a Fermi
frame, $X_a\equiv\partial_{\displaystyle{x^a}},a\in[0,3]$,
satisfies orthonormality
\begin{eqnarray}
g(X_a,X_b)|_C=\eta_{ab} \label{B05}
\end{eqnarray}
and parallel transport
\begin{eqnarray}
\nab{X_a}{X_b}|_C=0. \label{B10}
\end{eqnarray}
The $C$ subscript indicating evaluation at points on the curve -
the \emph{origin} of the spatial coordinates. Equation (\ref{B10})
follows from  orthonormality (\ref{B05}) and ${\mathcal
L}_{\displaystyle{X_a}}X_b=\nab{X_a}{X_b}-\nab{X_b}{X_a}=0$, since
on the curve
\begin{eqnarray*}
g(\nab{X_a}{X_b},X_c)&=-g(X_b,\nab{X_a}{X_c})\\
&=-g(X_b,\nab{X_c}{X_a})\\ &=g(X_a,\nab{X_c}{X_b})\\
&=g(X_a,\nab{X_b}{X_c})\\ &=-g(X_c,\nab{X_b}{X_a})\\
 &=-g(X_c,\nab{X_a}{X_b})
\end{eqnarray*}
implies $g(\nab{X_a}{X_b},X_c)=0$ for arbitrary $a,b,c\in[0,3]$.
\subsection{The evaluation of the connection near the world
line}
Any event near the world line of an observer not only defines its
Fermi coordinates but also a unique spacelike geodesic connecting
the observer with the event. This allows tensors to be parallel
transported to the world line in a special way and Taylor
expansion of tensor fields in the neighbourhood of $C$ is well
defined. A tensor $A|_U$ at spatial position corresponding to the
spacelike vector $U$ via the exponential map, can be expanded
\begin{eqnarray*}
A|_U=A|_C+\nab{U}{A}|_C+\ha\nab{U}{\nab{U}{A}}|_C+O(|U|^3)
\end{eqnarray*}
where $A|_C$ indicates its value on the world line. At arbitrary
time $x^0$ an arbitrary spacelike tangent vector, $U$, at $C(x^0)$
is mapped by the exponential map to a unique spacelike geodesic
$\cal U$. Let $\cal U$ have affine parameter $u$ so that
$U=\partial_u$. Then $\nab{U}{U}=0$ along $\cal U$ implies that
$\nab{U}{\nab{U}{U}}=0$ along $\cal U$. Let $V$ be another
arbitrary tangent vector at $C(x^0)$ then using
$\nab{U}{\nab{U}{U}}=\nab{V}{\nab{V}{V}}=0$ implies on setting
$Y=U+V$ that
\begin{eqnarray*}
\nab{Y}{\nab{Y}{Y}}=0=\nab{U}{\nab{U}{V}}+\nab{U}{\nab{V}{V}}
+\nab{V}{\nab{U}{U}}+\nab{V}{\nab{V}{U}}
\end{eqnarray*}
We also find that
\begin{eqnarray*}
\nab{Y}{Y}=0&=&\nab{U}{U}+\nab{U}{V}+\nab{V}{U}+\nab{V}{V}\\
 &=&2\nab{U}{V}
\end{eqnarray*}
confirming  (\ref{B10}). Using the above result, let $Z=U+V+W$
where $U,V,W$ are all arbitrary spacelike vectors in the tangent
space of $C(x^0)$ then
\begin{eqnarray*}
\nab{Z}{\nab{Z}{Z}}=0=3(\nab{U}{\nab{V}{W}}+\nab{V}{\nab{W}{U}}+\nab{W}{\nab{U}{V}}).
\end{eqnarray*}
Using this result and the definition of the Riemmann curvature
tensor in terms of arbitrary vector fields $E,F,G$,
\begin{eqnarray*}
\Riem(E,F,G,-)\equiv\nab{E}{\nab{F}{G}}-\nab{F}{\nab{E}{G}}-\nab{[E,F]}{G}
\end{eqnarray*}
where the last term vanishes when $[E,F]\equiv\Lie{E}{F}=0$ as is
the case here, implies
\begin{eqnarray*}
\Riem(U,V,W,-)+\Riem(U,W,V,-)\\=\nab{U}{\nab{V}{W}}-\nab{V}{\nab{U}{W}}
+\nab{U}{\nab{W}{V}}-\nab{W}{\nab{U}{V}}\\
=3\nab{U}{\nab{V}{W}}-\nab{U}{\nab{V}{W}}-\nab{V}{\nab{U}{W}}-\nab{W}{\nab{U}{V}}\\
=3\nab{U}{\nab{V}{W}}
\end{eqnarray*}
so in particular for $a,b\in[1,3]$ and $U$ spacelike
\begin{eqnarray*}
\nab{U}{\nab{X_a}{X_b}}|_C=\frac{1}{3}
\big(\Riem(U,X_a,X_b,-)|_C+\Riem(U,X_b,X_a,-)|_C\big).
\end{eqnarray*}
For the remaining cases, let $a\in[0,3]$ then
\begin{eqnarray}
\nab{X_0}{X_a}|_U=\nab{X_a}{X_0}|_U&=\nab{U}{\nab{X_0}{X_a}}|_C\nonumber\\
&=\Riem(U,X_0,X_a,-)|_C+\nab{X_0}{\nab{U}{X_a}}|_C\nonumber\\
&=\Riem(U,X_0,X_a,-)|_C. \label{A65}
\end{eqnarray}
In particular for $a=0$
\begin{eqnarray}
\nab{X_0}{X_0}|_U&=\Riem(U,X_0,X_0,-)|_C\nonumber\\
&=-\Riem(X_0,U,X_0,-)|_C. \label{A70}
\end{eqnarray}
It is instructive to compare (\ref{A70}) with the equation of
geodesic deviation. Let $Q$ be a geodesic vector field so that
$\nab{Q}{Q}=0$. Introduce the field $S$ orthogonal to $Q$ such
that $\Lie{S}{Q}=0$, then
\begin{eqnarray*}
\nab{S}{\nab{Q}{Q}}=0&=\nab{S}{\nab{Q}{Q}}-\nab{Q}{\nab{S}{Q}}+\nab{Q}{\nab{S}{Q}}\\
&=-\Riem(Q,S,Q,-)+\nab{Q}{\nab{S}{Q}}
\end{eqnarray*}
so that using $\Lie{S}{Q}=0$ gives
\begin{eqnarray*}
\nab{Q}{\nab{Q}{S}}=\Riem(Q,S,Q,-).
\end{eqnarray*}
Now $S$ represents the physical separation of two neighbouring
geodesics. On the other hand $U$ in (\ref{A70}) represents an
observer a fixed displacement from a geodesic and so must be
accelerating in an opposite sense to $S$ to maintain its position.

We require one further property regarding the orthonormality of
the Fermi frame. To $O(|U|)$:
\begin{eqnarray*}
g(X_a,X_b)|_U&=&g(X_a,X_b)|_C+g(\nab{U}{X_a},X_b)|_C+g(X_a,\nab{U}{X_b})|_C\\
&=&g(X_a,X_b)|_C.
\end{eqnarray*}
\section{A relativistic string in Fermi normal coordinates}
The above results suggest the definition of a new tensor
associated with the Fermi frame. Let general vectors $A,B$ be
decomposed into the time components ${\cal A,B}$ and 3 space
components ${\bf A,B}$ of the Fermi frame (e.g. ${\cal
A}=A^0X_0\equiv A^0\dot C,{\bf A}=A^aX_a$), then for spacelike
displacement $\bf R$ define the (2,1) \emph{Fermi} tensor, $F_{\bf
R}$ by
\begin{eqnarray*}
\mathrm{F}_{\bf R}(A,B,-)&\equiv&\Riem({\bf R},{\cal
A,B},-)\!+\!\Riem({\bf R},{\cal A},{\bf B},-)+\Riem({\bf R},{\bf
A},{\cal B},-)\\&&+\frac{1}{3}[\Riem({\bf R},{\bf
A,B},-)+\Riem({\bf R},{\bf B,A},-)]
\end{eqnarray*}
For example, with $a\in[0,3]$, the following term taken from
(\ref{RS}) expands as
\begin{eqnarray*}
\rho\nab{V}{V}|_{{\bf R}}=\rho\alpha\dot V^a
X_a+\rho\mathrm{F}_{\bf R}(V,V,-),
\end{eqnarray*}
where the first term is identical to that derived for flat
spacetime and the second term contains the curvature. The
definition of $F_{\bf R}$ yields
\begin{eqnarray*}
\mathrm{F}_{\bf R}(V,V,-)&=&(\alpha\dot T)^2\Riem({\bf R},\dot
C,\dot C,-)
 +2\alpha^2\dot T\Riem({\bf R},\dot C,\dot{\bf
 R},-)\\
 &&+\frac{2}{3}\alpha^2\Riem({\bf R},\dot{\bf R},\dot{\bf R},-).
\end{eqnarray*}
The Fermi tensor allows a succinct upgrade of the special
relativistic equations to Fermi relativistic equations. Using the
general relativistic equation (\ref{RS}), the flat spacetime
results in (\ref{S7}) and (\ref{S8}) and the definition of $F_{\bf
R}$ we obtain:
\begin{eqnarray}
0&=&\alpha\beta[\partial_\tau(\alpha\dot
T\rho_0)\partial_t+\partial_\tau(\alpha\rho_0\dot{\bf R})]
+\alpha\beta\left[\left(\frac{\beta p_0}{\dot T}\,\dot{\bf
R}\cdot{\bf R'}\right)'\partial_t+(\beta p_0{\bf
R}')'\right]\nonumber\\
 &&\!+\!\rho\mathrm{F}_{\bf R}(V,V,-)\!+\!p\mathrm{F}_{\bf R}(W,W,-)\!+\!\rho
 V\mathrm{F}_{\bf R}(W,V,\tilde W)\!+\!pW\mathrm{F}_{\bf R}(V,V,\tilde
 W),\nonumber\\
  \label{RSF}
\end{eqnarray}
which together with the substitutions: $V=\alpha(\dot T \dot
C+\dot{\bf R})$, $W=\beta(T'\dot C+{\bf R}')$, $T'={\bf \dot
R\cdot R'}/\dot T$, $\rho=\rho_0\beta$ and $p=p_0\alpha$, complete
the Fermi relativistic string.

We note that at non relativistic speeds the dominant curvature
term in $\mathrm{F}_{\bf R}(V,V,-)$  is $\Riem({\bf R},\dot C,\dot
C,-)$. It can be shown that this is the dominant term in
(\ref{RSF}) and the string equation becomes in the Newtonian limit
\begin{eqnarray*}
\rho_0\ddot{\bf R}={\bf N}'+\rho_0\Riem(\dot C,{\bf R},\dot C,-).
\end{eqnarray*}
using antisymmetry of $Riem$ in the first two slots. To order the
terms in magnitude note that $ p_0/\rho_0$ has dimensions
Force/(mass per unit length), or speed squared and may be treated
as $p_0/\rho_0\sim c_s^2$ where $c_s$ is the characteristic speed
of `sound' of the string. This may be taken in many cases to be
greater  than $v$ avoiding `shocks'. Coupling the next two most
dominant curvature terms to the Cosserat string then yields
\begin{eqnarray}
\rho_0\ddot{\bf R}-{\bf N}'&=&\rho_0\Riem(\dot C,{\bf R},\dot
C,-)\nonumber\\
 &&+\Pi_C\left[2\rho_0\Riem(\dot C,{\bf R},\dot{\bf R},-)
 +\frac{2}{3}\Riem({\bf R, R', N},-)\right]
\label{CossG}
\end{eqnarray}
where $\Pi_C$ is the projection onto the rest space of the
observer.

Equation (\ref{CossG}) has dimensions $ML^{-1}T^{-2}$. Using the
constitutive relation (\ref{Constitutive}) and
\begin{eqnarray*}
{\bf r}\equiv\frac{\bf R}{L_0},\quad
c_s\equiv\sqrt{\frac{E}{\rho_0}},\quad {\bf
n}\equiv(\nu-1)\frac{\bf r'}{\nu},
\end{eqnarray*}
(\ref{CossG}) can be written
\begin{eqnarray*}
\fl\ddot{\bf r}-c_s^2{\bf n}'=\Riem(\dot C,{\bf r},\dot
C,-)+\Pi_C\left[2\Riem(\dot C,{\bf r},\dot{\bf
r},-)+\frac{2c_s^2}{3}\Riem({\bf r,r',n,-})\right].\\
\end{eqnarray*}
Setting $c_s=1$ gives
\begin{eqnarray}
\ddot{\bf r}-{\bf n}'={\bf f}
\label{CossG2}
\end{eqnarray}
where
\begin{eqnarray}
{\bf f}\equiv\Riem(\dot C,{\bf r},\dot C,-)+\Pi_C\left[2\Riem(\dot
C,{\bf r},\dot{\bf r},-)+\frac{2}{3}\Riem({\bf r,r',n,-})\right].
\label{force}
\end{eqnarray}
This fixes the unit of length to that of the unstretched string,
$L_0$; the unit of speed to $c_s$; and the unit of time to
$L_0/c_s$.

In (\ref{CossG2}) the tidal force, $\Riem(\dot C,{\bf r},\dot
C,-)$, is clearly the dominant force. However, the other terms are
not completely negligible. Consider for example a string arranged
in a closed circular ring spinning about a normal axis through its
centre. Then this arrangement can respond so as to divert the ring
from a geodesic path via the second and third terms (the tidal
term cannot do this). There is a connection here with Dixon's
equations \cite{Dixon}  (see  also \cite{Moh} for the dynamics of
spinning particles in gravitational waves) as well as an analogy
with electromagnetic effects \cite{Cio}. The first term is
analogous to an electric field; the second term a magnetic field
due to the coupling between mass current and the field. The third
term, however, has no electromagnetic analogue and is generally
weaker than the magnetic term. The next two sections investigate
the dynamics of Cosserat string arranged in an axially flowing
ring- a lasso.
\section{Dynamics of a Cosserat string}
\label{six}
This section  seeks solutions to
\begin{eqnarray}
\ddot {\bf r}={\bf n}'+{\bf f} \label{main}
\end{eqnarray}
where
\begin{eqnarray}
{\bf n}=(|{\bf r}'|-1)\frac{{\bf r}'}{|{\bf r}'|} \label{nmain}
\end{eqnarray}
and $\dot{(\;)}\equiv\partial_\tau,(\;)'\equiv\partial_\sigma$. In
the absence of an external force, ${\bf f}=0$ a circular axially
flowing solution given in a Cartesian frame $\{{\bf i,j,k}\}$ is
\begin{eqnarray}
{\bf r}_0(\sigma,\tau)&=\frac{\nu}{2\pi}\cos(2\pi (\sigma+\mu
\tau)){\bf i}+\frac{\nu}{2\pi}\sin(2\pi (\sigma+\mu \tau)){\bf j}
\label{zerosol}
\end{eqnarray}
$\mu$ is the rotation frequency  related to the constant strain,
$\nu\equiv|{\bf r}_0'|$ by
\begin{eqnarray}
\mu \equiv\sqrt{\frac{\nu-1}{\nu}}
\end{eqnarray}
Defining an orthonormal frame rotating with each material point of
the string by

\begin{eqnarray}
\left[
\begin{array}{c}
 {\bf U}_1\\
{\bf U}_2\\
 {\bf U}_3\\
\end{array}
\right] \equiv\left[\begin{array}{ccc} \cos\theta & \sin\theta &
0\\ -\sin\theta & \cos\theta & 0\\ 0 & 0 & 1\\
        \end{array}
\right]\left[\begin{array}{c} {\bf i} \\ {\bf j}\\{\bf k}\\
\end{array}\right]
\label{UfromCart}
\end{eqnarray}
where
\begin{eqnarray}
\theta\equiv 2\pi (\sigma+\mu \tau)
\end{eqnarray}
solution (\ref{zerosol}) can be written
\begin{eqnarray}
{\bf r}_0(\sigma,\tau)&=\frac{\nu}{2\pi}{\bf U}_1.
\label{Zerothsol}
\end{eqnarray}
Perturbing about this result by writing for some small parameter,
$\epsilon$
\begin{eqnarray}
{\bf r}={\bf r}_0+\epsilon{\bf r}_1+O(\epsilon^2)
\end{eqnarray}
and demanding that ${\bf f}$ is 
\begin{eqnarray}
{\bf f}=O(\epsilon)
\end{eqnarray}
then inserting into (\ref{main}) and (\ref{nmain}) gives to
$O(\epsilon)$ the equations for ${\bf r}_1$

\begin{eqnarray}
&\left(\ddot\eta-\mu^2\eta''+2\pi(1+\mu^2)\xi'-4\pi\mu\dot\xi+4\pi^2(1-\mu^2)\eta\right){\bf
U}_1\nonumber\\
&+\left(\ddot\xi-\xi''-2\pi(1+\mu^2)\eta'+4\pi\mu\dot\eta\right){\bf
U}_2+(\ddot\zeta-\mu^2\zeta''){\bf U}_3={\bf f}
 \label{FirstOrder}
\end{eqnarray}
where
\begin{eqnarray}
{\bf r}_1=\eta{\bf U}_1+\xi{\bf U}_2+\zeta{\bf U}_3
\end{eqnarray}
The solution to the homogeneous equation, with ${\bf f}=0$ is
\begin{eqnarray}
{\bf r}_1=&\left[
\begin{array}{c}
\eta\\ \xi\\ \zeta\\ \end{array} \right]
 =\left[
\begin{array}{c}
A\cos(\omega_0\tau)+B\sin(\omega_0\tau)+C\\
\frac{4\pi\mu}{\omega_0}\left(-A\sin(\omega_0 \tau)+B\cos(\omega_0
\tau)\right)+\pi\frac{1-\mu_0^2}{\mu_0}C\tau+D\\ 0\\
\end{array}
\right]\nonumber\\ &+ \sum_{J=1}^2\left[
\begin{array}{c}
a_{J1}\alpha_{J1}\cos(\omega_{J1}\tau+2\pi
\sigma)+b_{J1}\alpha_{J1}\sin(\omega_{J1}\tau+2\pi \sigma)\\
a_{J1}\sin(\omega_{J1}\tau+2\pi
\sigma)-b_{J1}\cos(\omega_{J1}\tau+2\pi \sigma)\\ 0\\
\end{array}
\right]\nonumber\\ &+\sum_{k=2}^\infty\sum_{J=1}^{4}\left[
\begin{array}{c}
a_{Jk}\alpha_{Jk}\cos(\omega_{Jk}\tau+2\pi k
\sigma)+b_{Jk}\alpha_{Jk}\sin(\omega_{Jk}\tau+2\pi k \sigma)\\
a_{Jk}\sin(\omega_{Jk}\tau+2\pi k
\sigma)-b_{Jk}\cos(\omega_{Jk}\tau+2\pi k \sigma)\\0\\
\end{array}\right]\nonumber\\
&+\sum_{k=0}^\infty\left[
\begin{array}{c}
0\\0\\ c_k\cos(2\pi k(\mu\tau+\sigma))+d_k\sin(2\pi
k(\mu\tau+\sigma))\\
\end{array}
\right] \label{sol}
\end{eqnarray}
where $A,B,C,D,a_{Jk},b_{Jk},c_k,d_k$ are arbitrary constants,
\begin{eqnarray}
\alpha_{Jk}\equiv
\frac{-\omega_{Jk}^2+4\pi^2k^2}{(-4\pi^2k(1+\mu^2)+4\pi\omega_{Jk}\mu)},
\label{10_EigenRatio}
\end{eqnarray}
\begin{eqnarray*}
\omega_0&=2\pi\sqrt{1+3\mu^2},\\
 \omega_{11}&=-2\pi(\mu-\sqrt{2(1+\mu^2)}),\\
\omega_{21}&=-2\pi(\mu+\sqrt{2(1+\mu^2)}),
\end{eqnarray*}
and, for each $k\geq 2$ and $J\in[1,4],\, \omega_{Jk}$ are the 4
solutions to the following equation in $x_k$
\begin{eqnarray}
\fl(2\pi\mu k-x_k)[x_k^3+2k\pi\mu
x_k^2-4\pi^2(3\mu^2+k^2+1)x_k+8k\pi^3\mu(\mu^2-k^2+3)]=0.
\label{Det0}
\end{eqnarray}
Solution, (\ref{sol}), satisfies the following restriction on the
centre of mass:
\begin{eqnarray*}
\int_0^1{\bf r}\de\sigma=0.
\end{eqnarray*}
When $k=0$ the term linear in $\tau$ can be interpreted as changes
to the angular velocity of the ring. The ${\bf U}_3$ motion,
governed by $\zeta$, is a string in static equilibrium and in the
special case of $k=2$ can be interpreted as a fixed rigid body
rotation of the ring about a bisecting axis.

\begin{figure}[t]
\unitlength1cm
\begin{minipage}[t]{7.0cm}
\includegraphics[width=7cm]{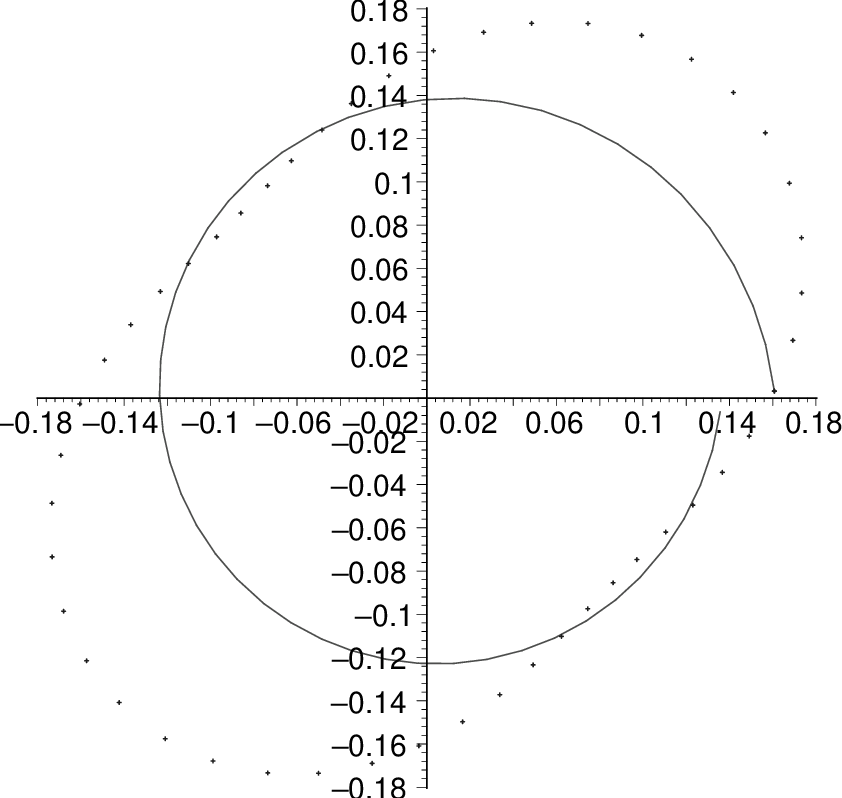}
\caption{The $k=2$ motion for $\omega_{12}$} \label{10_1_NM_Omega}
\end{minipage}
\hfill
\begin{minipage}[t]{7.0cm}
\includegraphics[width=7cm]{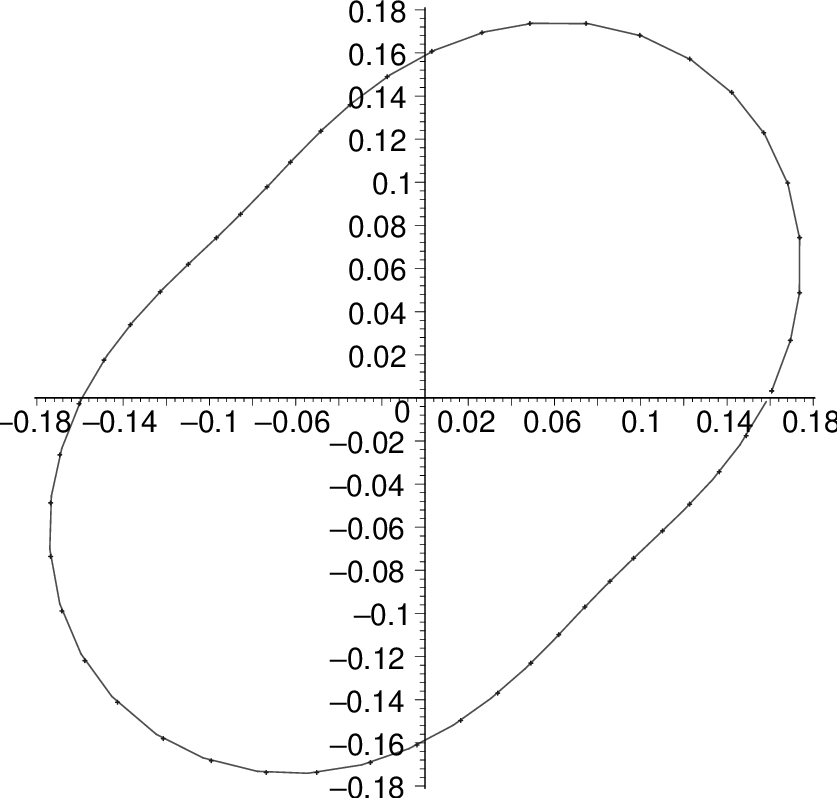}
\caption{The $k=2$ motion for $\omega_{22}$} \label{10_2_NM_Omega}
\end{minipage}
\end{figure}
\begin{figure}[t]
\unitlength1cm
\begin{minipage}[t]{7.0cm}
\includegraphics[width=7cm]{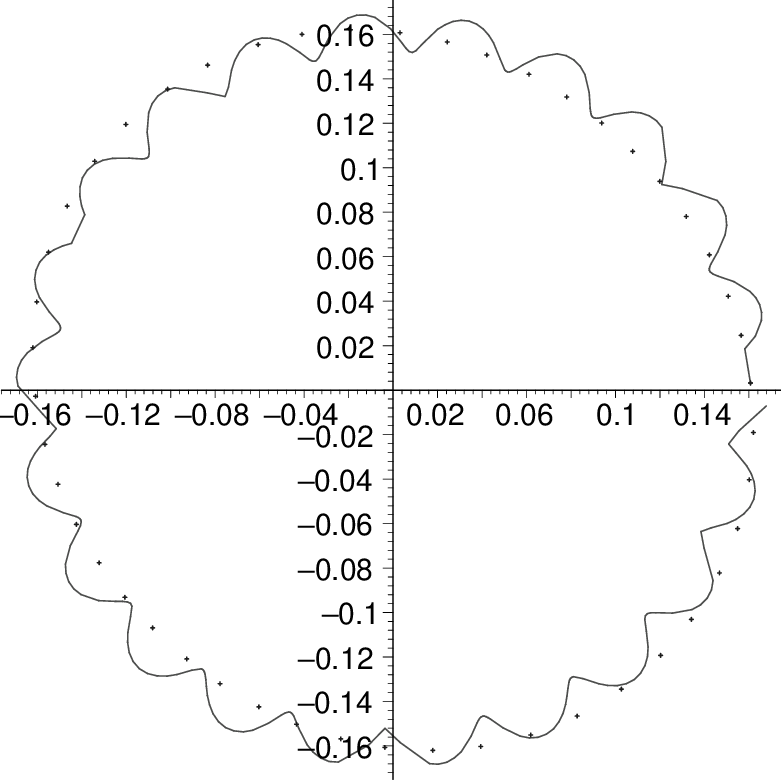}
\caption{The $k=2$ motion for $\omega_{32}$} \label{10_3_NM_Omega}
\end{minipage}
\hfill
\begin{minipage}[t]{7.0cm}
\includegraphics[width=7cm]{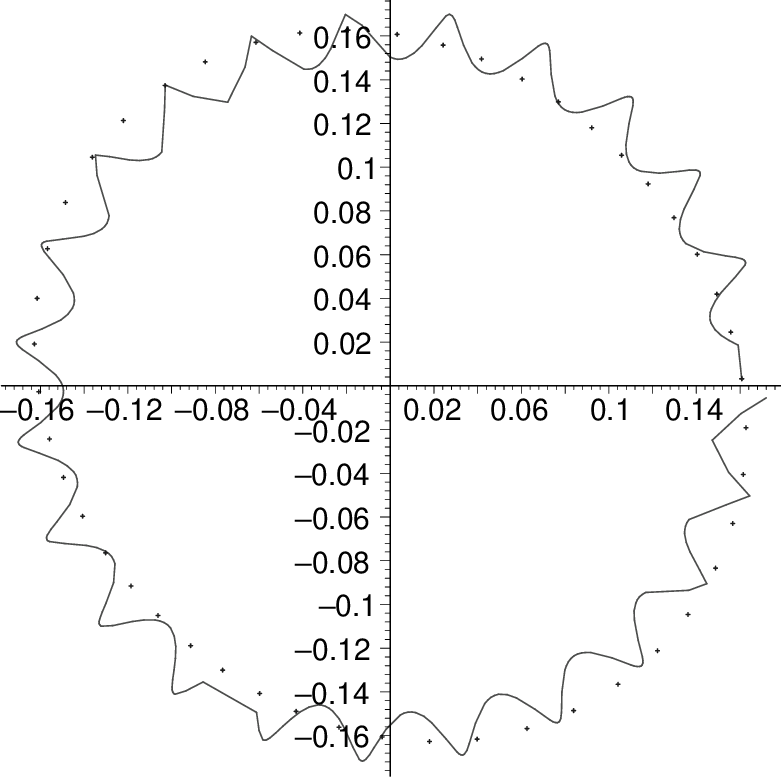}
\caption{The $k=2$ motion for $\omega_{42}$} \label{10_4_NM_Omega}
\end{minipage}
\end{figure}

 Apart from the zeroth and first modes there are 4 characteristic
frequencies associated with each mode shape governed by $k$. For
$k=2$, figures (\ref{10_1_NM_Omega}) to (\ref{10_4_NM_Omega}) show
the mode shape at $\tau=0$ (dotted line) and the motion of the
material point $\sigma=0$ (continuous line) over a time period
just short of one period for a stretch of $\nu_0=1.01$. Figure
(\ref{10_2_NM_Omega}) clearly shows  an axially flowing `Healey'
loop \cite{Healey}.  Also of interest is the motion corresponding
to the lowest frequency which nearly approximates rigid body
rotation (exact rotation is characterized by circular motion). For
higher stretch and modes the picture is essentially the same.
\section{Motion of the ring in a plane monochromatic gravitational wave}
In this section, using both Fermi coordinates and a related system
(\emph{plane} normal coordinates), the relative acceleration
associated with a monochromatic plane gravitational wave is
derived on a volume of space surrounding a line of observers. The
wave frequencies that parametrically excite the ring of section
(\ref{six}) are determined and non geodesic motion due to 'spin'
coupling is demonstrated.

Writing
\begin{eqnarray}
g\equiv g_M+h
\end{eqnarray}
where, with $e^0\equiv c\de t,e^1\equiv \de x,e^2\equiv \de
y,e^3\equiv \de z$, $g_M$ is the Minkowski metric
\begin{eqnarray}
g_M\equiv -e^0\otimes e^0+e^1\otimes e^1+e^2\otimes e^2+e^3\otimes
e^3,
\end{eqnarray}
$c$ is the speed of light and $h$ is defined
\begin{eqnarray}
h\equiv \epsilon\kappa
(2xy+x^2-y^2)\sin\left(\frac{2\pi(z-ct)}{\lambda}\right)(e^0-e^3)\otimes(e^0-e^3).
\label{h}
\end{eqnarray}
The  metric $g$ is an {\em exact} vacuum solution of Einstein's
equations for arbitrary dimensionless constant $\epsilon$,
constant $\kappa$ with dimensions $L^{-2}$ and represents a plane
monochromatic gravitational wave \cite{SW},\cite{MTW}. Now the
frame $\{e^a\}$ defines the metric dual frame $\{X_a\}$ which
coincides with a Fermi frame $\{{\bf W}_a=X_a|_C,a=[1,3]\}$ for an
observer defined by
\begin{eqnarray*}
\tau\mapsto C(\tau)\equiv(t=\tau,x=0,y=0,z=0)
\end{eqnarray*}
so that
\begin{eqnarray*}
\dot C\equiv \partial_\tau=\partial_t
\end{eqnarray*}
and the tidal force at a position ${\bf r}\equiv r^aW_a$ relative
to $C$ is given by
\begin{eqnarray}
\Riem(\dot C,{\bf r},\dot C,-)=\epsilon c^2\kappa
\sin\left(\frac{-2\pi c\tau}{\lambda}\right)[(r^1+r^2){\bf
W}_1+(r^1-r^2){\bf W}_2]. \label{12_E_GWave}
\end{eqnarray}
valid for small $r^3\ll\lambda$. The tidal force can be
interpreted \cite{SW} by its effects in a tangent plane, $W$,
normal to the propagation direction
$Y\equiv\frac{1}{c}\partial_t+\partial_z$, lying on the
intersection of the rest space of a geodesic observer and the
level hypersurfaces of $ct-z$. For the observer $C$ the plane $W$
is spanned by ${\bf W}_1$ and ${\bf W}_2$ at $x=y=z=0$, and can be
interpreted as the 2-dimensional wavefront associated with $C$.
For short wave lengths, $r^3>\lambda$ the tidal force
(\ref{12_E_GWave}) is only reliable when restricted to $W$ since
it is only accurate to first order in distance from the observer.
To overcome this, consider a one parameter family of observers
given by the mapping, $S$ defined by
\begin{eqnarray*}
(\tau,\upsilon)\mapsto S(\tau,\upsilon)\equiv
(t=\tau,x=0,y=0,z=\upsilon)
\end{eqnarray*}
each with 2-dimensional wavefront $W^\upsilon$ spanned by
$\partial_x,\partial_y$ on $S(\tau,\upsilon)$. In this way it is
possible to discuss relative accelerations induced by the wave in
a volume of space defined by a sequence of tangent planes
$W^\upsilon$ by
\begin{eqnarray}
\fl\Riem(\dot S,{\bf r},\dot S,-)=\epsilon c^2\kappa
\sin\left(\frac{2\pi(\upsilon-c\tau)}{\lambda}\right)[(r^1+r^2){\bf
W}_1(\upsilon)+(r^1-r^2){\bf W}_2(\upsilon)]. \label{Tidal2}
\end{eqnarray}
where $\dot S\equiv\partial_t$ on $S(\tau,\upsilon)$ and the frame
basis , $\{{\bf W}_1,{\bf W}_2\}$ depends both on $\tau$ and
$\upsilon$. If, as is the case here, $S(\tau,\upsilon)$ is aligned
along a spacelike geodesic
$(\nab{\partial_\upsilon}{\partial_\upsilon}=0)$ then it is
possible to associate the vector $r^3{\bf W}_3$ via the
exponential map at time $\tau$ to an observer in $S$ by
$S(\tau,\upsilon)=\exp_\tau(r^3{\bf W}_3)$ and hence
$\upsilon=r^3$.

To put all this in context, consider the following coordinate
systems: Normal (or Gaussian) coordinates, which are 4 coordinates
defined in a tangent space at an event in spacetime
\cite{Nesterov}; Fermi normal coordinates which are 3 coordinates
defined at a series of one parameter
 volume tangent spaces; and, for want of a better name, {\it Plane} normal
coordinates which are 2 coordinates defined at a series of two
parameter plane tangent spaces representing a string of observers.
The latter is what we have here. We might term these $(0,4),
(1,3)$ and $(2,2)$ normal coordinate systems respectively. So, for
example, $(4,0)$ normal coordinates are ordinary coordinates and
$(3,1)$ normal coordinates presumably have application for world
volume corresponding to a surface of observers.

We now inspect the force on the unperturbed spinning ring
(\ref{Zerothsol}) with observer $C$ located at the circle centre
and assume $\epsilon\ll 1$ so that a force ${\bf f}$ can be
equated with a perturbation to the static equilibrium solution
(\ref{Zerothsol}) so that (\ref{FirstOrder}) can be applied. On
the ring
\begin{eqnarray}
r^1=r_0\cos\theta,\nonumber\\ r^2=r_0\sin\theta,\nonumber\\ r^3=0,
\label{InPlane}
\end{eqnarray}
where $r_0\equiv 1/(2\pi(1-\mu^2))$, $\theta\equiv
2\pi(\sigma+\mu\tau)$ and the Fermi basis ${\bf W}_a, a\in[1,3]$
replaces the Cartesian basis  given in terms of the $\{{\bf
U}_a\}$ basis by inverting (\ref{UfromCart})
\begin{eqnarray*}
\left[
\begin{array}{c}{\bf W}_1 \\ {\bf W}_2\\{\bf W}_3\\
\end{array}
\right] \equiv\left[\begin{array}{ccc} \cos\theta & -\sin\theta &
0\\ \sin\theta & \cos\theta & 0\\ 0 & 0 & 1\\
        \end{array}
\right]\left[ \begin{array}{c}
 {\bf U}_1\\
{\bf U}_2\\
 {\bf U}_3\\
\end{array}\right]
\end{eqnarray*}
Substituting into (\ref{12_E_GWave}) gives the rhs of
(\ref{FirstOrder}):
\begin{eqnarray*}
{\bf f}&=&r_0c^2\kappa \sin\left(\frac{-2\pi
c\tau}{\lambda}\right) [(\cos\theta+\sin\theta){\bf W}_1
+(\cos\theta-\sin\theta){\bf W}_2]\\
 &=&\ha r_0c^2\kappa\bigg\{[\sin(\phi_+)-\sin(\phi_-)-\cos(\phi_+)+\cos(\phi_-) ]{\bf
U}_1\\
 &&+[\sin(\phi_+)-\sin(\phi_-)+\cos(\phi_+)-\cos(\phi_-)]{\bf U}_2\bigg\} \label{f}
\end{eqnarray*}
where
\begin{eqnarray*}
\phi_{\pm}\equiv 2\theta\pm\frac{2\pi
c\tau}{\lambda}\equiv4\pi\sigma+(4\pi\mu\pm 2\pi c/\lambda)\tau
\end{eqnarray*}
and light speed, $c\equiv c/c_s$, is a dimensionless constant. We
expect resonant effects to take place when the forcing frequencies
match the resonant frequencies
\begin{eqnarray*}
\omega_{Jk}\tau+2\pi k\sigma=\phi_{\pm}\equiv(4\pi\mu\pm 2\pi
c/\lambda)\tau+4\pi\sigma
\end{eqnarray*}
This occurs when $k=2$ and
\begin{eqnarray}
\omega_{J2}=4\pi\mu\pm 2\pi c/\lambda. \label{resonance}
\end{eqnarray}
Substituting $k=2$ and (\ref{resonance}) into (\ref{Det0}) gives a
cubic equation for the angular frequency, $\mu$. For example,
consider the 'Healey' solution to (\ref{Det0}) for $k=2$ and using
the label $J=2$
\begin{eqnarray*}
\omega_{22}=4\pi\mu.
\end{eqnarray*}
This implies that resonance occurs as the frequency, $f\equiv
c/\lambda$, of the gravitational wave approaches zero. Recall that
a Healey loop is in static equilibrium and excitation of
$\omega_{22}$ would result in a gradual distortion of the ring.
However, a wave with $f=0$ is not a wave at all so this result may
be discounted. The remaining frequencies are dependent on
$\mu\equiv\sqrt{1-1/\nu}$. Consider a particular strain,
$\nu=1.01$ implying $\mu=0.0995$ and solutions to (\ref{Det0}) of
\begin{eqnarray*}
\omega_{12} =-0.2458,\quad \omega_{22} =1.2504,\quad \omega_{32}
=13.6067,\quad \omega_{42} =-14.6113.
\end{eqnarray*}
Then corresponding wave frequencies $f_J$ associated with
resonance in this example are
\begin{eqnarray*}
f_1=\pm 0.238,\quad f_2=0,\quad f_3=\pm 1.967,\quad f_4=\pm 2.524.
\end{eqnarray*}
It is not clear which mode of vibration would be easiest to
detect.

Assume that there exists a ring corresponding to the above
example. The question arises as to how the ring could be altered
without effecting the resonant frequencies, $f_J$. Since $f_J$ are
only dependent on time which is measured in units of
characteristic time $L_0/c_s$, there is a class of rings with
equal characteristic time which will resonate with a particular
wave of given frequency. For example one may keep the
characteristic time constant by doubling length $L_0$ and also
doubling $c_s=\sqrt{E/\rho_0}$ by halving the cross sectional
diameter of the string, since the mass per unit length $\rho_0$ is
proportional to area, $A$. Hence, there is no theoretical limit to
the size of the ring. However, practical considerations may decree
that the ring may be of comparable size to the wavelength,
$L_0\approx\lambda$ and it is desirable to know how such a
detector may respond when not correctly aligned in the wave front.
To do this (\ref{Tidal2}) is used in place of (\ref{12_E_GWave})
and the extreme example where the normal to the ring is
perpendicular to the propagation direction is  now discussed. With
the ring lying in the $xz$ plane

\begin{eqnarray}
\left[
\begin{array}{c}{\bf W}_1 \\ {\bf W}_2\\{\bf W}_3\\
\end{array}
\right] \equiv\left[\begin{array}{ccc} \cos\theta & -\sin\theta &
0\\ 0 & 0 & -1\\ \sin\theta & \cos\theta & 0\\
        \end{array}
\right]\left[ \begin{array}{c}
 {\bf U}_1\\
{\bf U}_2\\
 {\bf U}_3\\
\end{array}\right]
\label{Frame2}
\end{eqnarray}
and
\begin{eqnarray}
r^1=r_0\cos\theta,\nonumber\\ r^2=0,\nonumber\\
r^3=\upsilon=r_0\sin\theta. \label{Plane2}
\end{eqnarray}
As with (\ref{InPlane}) we assume that the ring configuration
(\ref{Zerothsol}), when written in the Fermi frame is a solution
to Cosserat's equations to $O(\epsilon)$, even for short
wavelength $\lambda$. Substitution of (\ref{Frame2}) and
(\ref{Plane2}) into the tidal tensor (\ref{Tidal2}) gives
\begin{eqnarray*}
\fl\Riem(\dot S,{\bf R},\dot S,-)&=&\epsilon c^2\kappa
\sin\left(\frac{2\pi(r_0\sin\theta-c\tau)}{\lambda}\right)
r_0\cos\theta{\bf W}_1\\ &=&\epsilon r_0c^2\kappa
\sin\left(\frac{2\pi(r_0\sin\theta-c\tau)}{\lambda}\right)
\cos\theta(\cos\theta{\bf U}_1-\sin\theta{\bf U}_2).
\end{eqnarray*}
The case $r_0\rightarrow 0$ results in terms cosine or sine of
$(2\theta\pm 2\pi f\tau)$ which we can use to extract from
$\Riem(\dot S,{\bf R},\dot S,-)$ coefficients of similar terms
when $L_0\approx 2\pi r_0\in[0,\lambda]$ by inspecting the change
in the following integrals
\begin{eqnarray*}
\int_0^1\int_0^b\Riem(\dot S,{\bf R},\dot S,-)\sin(2\theta+ 2\pi
f\tau)\de\tau\de\sigma,\\ \int_0^1\int_0^b\Riem(\dot S,{\bf
R},\dot S,-)\cos(2\theta+ 2\pi f\tau)\de\tau\de\sigma,
\end{eqnarray*}
where $b\equiv 1/(2\mu+f)$ and $f\equiv\pm c/\lambda$. Numerical
evaluation indicates little change for $L_0$ in the range given,
implying that resonant effects are not destroyed when the ring is
not aligned perfectly in the wave front.

To examine the motion of the centre of mass (CM), note that by the
linearity of the tidal tensor and the symmetry of the ring that
the acceleration of the CM vanishes
\begin{eqnarray*}
\int_0^1\Riem(\dot C,{\bf R},\dot C,-)\de\sigma=0.
\end{eqnarray*}
However the term
\begin{eqnarray*}
\epsilon\ddot{\bf R}_{CM}\equiv 2\int_0^1\Riem(\dot C,{\bf R},\dot
{\bf R},-)\de\sigma.
\end{eqnarray*}
does not vanish in general. Using the ring configuration
(\ref{Frame2}) and (\ref{Plane2}) implying
\begin{eqnarray*}
\dot{\bf R}=2\pi \mu r_0(-\sin\theta{\bf W}_1+\cos\theta{\bf
W}_3).
\end{eqnarray*}
Specializing to long wavelengths ($r^3\ll\lambda$), gives
\begin{eqnarray*}
\ddot{\bf R}_{CM}=2\pi r_0^2\kappa c\mu \sin(2\pi f\tau)({\bf
W_1+W_2})
\end{eqnarray*}
and hence the centre of mass ${\bf R}_{CM}$ will oscillate in
simple harmonic motion about the geodesic observer with amplitude
proportional to the spin frequency, $\mu$.  Finally, since ${\bf
n}=(1-1/\nu){\bf r}'\equiv \mu^2{\bf r}'$ it is easily verified
that $\int_0^1 \Riem({\bf r,r',n,-})\de\sigma$, vanishes in both
orientations and thus does not contribute to the $CM$ motion.
\section{Conclusion}
This paper has derived a general relativistic string, its
reduction in the special case of flat spacetime, its form in a
local Fermi frame  and, for non relativistic speeds a formulation
as a Cosserat string plus three gravitational terms. By
perturbative analysis the normal modes associated with a Cosserat
string arranged as an axially flowing ring were investigated. This
allowed analysis of the effects associated with a ring in a plane
monochromatic gravitational wave, in particular the excitation of
the ring's natural frequencies and the oscillation of its centre
of mass about geodesic motion.

There remain, of course, significant practical obstacles to the
construction of a lasso antenna, not least how to convert string
motion to data  with sufficient sensitivity to extract those
resonances associated with gravitation waves from more dominant
Newtonian effects. Piezoelectric material or optical fibre may
have properties that can be exploited in this respect.


\section*{References}
\begin{thebibliography}{}
\bibitem{MTW}Misner C W, Thorne K, Wheeler A 1973 {\it Gravitation} San
Fransisco. Freeman
\bibitem{Finn}Finn 1996, {\it Gravitational radiation sources for
acoustic detectors} preprint gr-qc/9609027
\bibitem{Ant} Antman S 1991 {\it Non-linear problems in elasticity}
Applied Mathematical Sciences 107  N.Y. Springer-Verlag.es
\bibitem{TuckerWang} Tucker R W, Wang C 1999 {\it Journal of Sound and Vibration} {\bf
224} (1) 123-165
\bibitem{Tucker} Benn I M, Tucker R W 1987 {\it An
Introduction to Spinors and Geometry with Applications in Physics}
Adam Hilger, Bristol and New York
\bibitem{Nesterov} Nesterov A 1999 {\it Class. Quantum Grav.} {\bf 16}(2) 465-477, 1999.
\bibitem{Dixon} Dixon W G 1969 {\it Proc. R. Soc.} A {\bf 314} 499
\bibitem{Moh} Mohseni M, Tucker R W, Wang C 2001 {\it Class. Quantum
Grav.}{\bf 18} 1-11
\bibitem{Cio} Ciofolini I, Wheeler A {\it Gravitation and Inertia} Princeton series in Physics
\bibitem{Healey} Healey T 1990 {\it Quarterly of Applied Mathematics}
{\bf 48} 679-698
\bibitem{SW}Sachs R K, Wu H, {\em General Relativity for
Mathematicians} New York. Springer. c1977

\end{thebibliography}
\end{document}